\documentstyle[epsf,epsfig,12pt]{article}

\newlength{\dinwidth}
\newlength{\dinmargin}
\newcommand{\ba}{\begin{array}}
\newcommand{\ea}{\end{array}}
\newcommand{\be}{\begin{equation}}
\newcommand{\ee}{\end{equation}}
\newcommand{\bea}{\begin{eqnarray}}
\newcommand{\eea}{\end{eqnarray}}
\newcommand{\beas}{\begin{eqnarray*}}
\newcommand{\eeas}{\end{eqnarray*}}

\def\to{\rightarrow}
\def\bee{\begin{eqnarray}}
\def\eee{\end{eqnarray}}
\def\be{\begin{equation}}
\def\ee{\end{equation}}

\font\cmss = cmss12

\def\laplace{{\kern1pt\vbox{\hrule height 1.2pt\hbox{\vrule width
1.2pt\hskip
  3pt\vbox{\vskip 6pt}\hskip 3pt\vrule width 0.6pt}\hrule height 0.6pt}
  \kern1pt}}
\def\scriptlap{{\kern1pt\vbox{\hrule height 0.8pt\hbox{\vrule width
0.8pt
  \hskip2pt\vbox{\vskip 4pt}\hskip 2pt\vrule width 0.4pt}\hrule height
0.4pt}
  \kern1pt}}

\def\roughly#1{\raise.3ex\hbox{$#1$\kern-.75em\lower1ex\hbox{$\sim$}}}

\def\sunk{{\hbox{\cmss SU($n|k$)}}}
\def\sunn{{\hbox{\cmss SU($n|n$)}}}
\def\sun0{{\hbox{\cmss SU($n|0$)}}}
\def\su-0{{\hbox{\cmss SU($n-k|0$)}}}

\def\u1{{\hbox{\cmss U(1)}}}
\def\qcpt{Q$\chi$PT}

\begin{document}
\thispagestyle{empty}
\addtocounter{page}{-1}
\begin{flushright}
CLNS 97/1506\\
IASSNS-HEP 97-89\\
SNUTP 97-102\\
{\tt hep-ph/9708432}\\
\end{flushright}
\vspace*{1cm}
\centerline{\Large \bf Quenched and Partially Quenched}
\vskip0.4cm
\centerline{\Large \bf Chiral Perturbation Theory}
\vskip0.4cm
\centerline{\Large \bf for}
\vskip0.4cm
\centerline{\Large \bf Vector and Tensor Mesons
\footnote{
Work supported in part by NSF Grant, NSF-KOSEF
Bilateral Grant, KOSEF Purpose-Oriented Research Grant 94-1400-04-01-3
and SRC-Program, Ministry of Education Grant BSRI 97-2410, the Monell
Foundation and the Seoam Foundation Fellowships.}}
\vspace*{1.2cm}
\centerline{\large\bf Chi-Keung Chow${}^a$ and Soo-Jong Rey${}^{b,c}$}
\vspace*{0.6cm}
\centerline{\large\it Newman Laboratory for Nuclear Studies}
\vskip0.1cm
\centerline{\large\it Cornell University, Ithaca NY 14853 USA${}^a$}
\vskip0.3cm
\centerline{\large\it School of Natural Sciences, Institute for Advanced
Study}
\vskip0.1cm
\centerline{\large\it Olden Lane, Princeton NJ 08540 USA${}^b$}
\vskip0.3cm
\centerline{\large\it Physics Department, Seoul National University,
Seoul 151-742 KOREA${}^c$}
\vspace*{0.6cm}
\centerline{\large\tt ckchow@mail.lns.cornell.edu,
sjrey@phya.snu.ac.kr}
\vspace*{1cm}
\centerline{\large\bf abstract}
\vskip0.5cm
Quenched and partially quenched chiral perturbation theory for vector
mesons is developed and is used to extract chiral loop correction to the
$\rho$ meson mass.
Connections to fully quenched and totally unquenched chiral perturbation
theory results are discussed.
It is also shown that (partially) quenched perturbation theory for
tensor mesons can be formulated analogously, and the chiral corrections
for tensor meson masses are directly proportional to their counterparts
in the vector meson sector. Utilizing this observation and non-relativistic
quark model, we point out that mass difference 
$(m_{a_2} - {3 \over 2} m_\rho)$ is ``quenching-insensitive'' in large-$N_c$
limit. This quantity
may be used for normalization of mass scale in lattice QCD calculations.
\vspace*{1.1cm}


\setlength{\baselineskip}{18pt}
\setlength{\parskip}{12pt}
\newpage

Lattice QCD simulations have reached such an impressive stage of high
precision that an accuracy up to a few percent error is expected to
be within a reach. Amongst such results are spectroscopy
of ground state hadrons including pseudo-Goldstone mesons, baryons
as well as some of the vector mesons~\cite{spec1, spec2, spec3, spec4, 
spec5, spec6}\footnote{for up-to-date review, see~\cite{specrev}.}.
More recently, the precision test of hadron spectroscopy on a lattice
has
gone beyond the ground state hadrons and has been extended, for example,
to tensor or exotic mesons~\cite{orbitaldata1, orbitaldata2, orbitaldata3}. 
The tensor mesons
are of particular interest since they are relatively clean states
experimentally, much more so than scalar mesons. As such, one might
hope that tensor mesons provide a good check-point on accuracy of
lattice QCD simulations. Technically, the difficulty has been that
gauge invariant lattice interpolating operator of tensor mesons is
non-local. Dynamial evolution of such operator is far more computer-time
consuming than low-lying mesons and baryons that require only local
operators, hence, limits accuracy of lattice data.
With the advent of recent results~\cite{orbitaldata2, 
orbitaldata3},
however, precision spectroscopy of the tensor mesons should become
possible in the foreseeable future.
Such an extension should be of importance in order to
gain a more complete insight and better understanding of the
nonperturbative aspects of QCD.

Simulation of full QCD on a lattice has turned out to be both time
consuming and costly.
Most of the simulation time is to calculate changes in the determinant
of the Dirac operator of light quarks. Because of this reason, so far,
lattice QCD calculations on a large-scale volume have been mainly
limited
to quenched (valence-quark) approximation in which gauge field
configurations are summed up with the determinant of
$N_F$ light quarks
${\rm det}^{N_F} ({\cal D} \hskip-0.22cm / (A) + m)$
is replaced by unity or, equivalently, $N_F \rightarrow 0$.
While the approximation is well suited when quark masses
are heavy enough~\cite{politzer}, 
extrapolation of the masses to physical, light quark
masses might results in potentially significant errors. It is therefore
necessary to understand how reliable the quenched approximation is and
where it begins to break down.
Indeed, Sharpe~\cite{S1, S2}
and Bernard and Golterman~\cite{BG1, BG2}
have pointed out that the quenched
approximation leads to sizable errors~\footnote{
Some relevant recent investigation includes~\cite{gupta, 
kimsinclair,michael}.}. 
If this is the case\footnote{For up-to-date review, see~\cite{
sharpereview}.}, 
then one needs to understand
better the errors incurred by quenched approximation before a reliable
hadron spectrum is extracted.

In order to study the error introduced by the quenched approximation
systematically,  Sharpe has developed quenched chiral
perturbation theory (\qcpt) \cite{S1,S2},
which is subsequently developed further by Bernard and
Golterman~\cite{BG1}~\footnote{For a pedagogical review, see \cite{
Greview}.}.
In order to cancel the effect of internal quark loops, following
the method proposed by Morel~\cite{morel}, Bernard and Golterman have
introduced the ghost quarks, which have the same masses as the
standard quarks, but with opposite statistics.
Diagrams with quark loops are thus cancelled by analogous diagrams with
ghost loops.
The quenched approximation is achieved by introducing a ghost quark to
every standard quark in the Lagrangian.
Just as standard chiral perturbation theory ($\chi$PT) is the low energy
effective theory of QCD, \qcpt~ 
is the low energy effective theory of this new
theory of quarks and ghosts.
One can then estimate the error introduced by the quenched approximation
by calculating the non-analytic contributions from chiral loops, and
compare them
with their counterparts in standard $\chi$PT.
If the structure of the chiral non-analyticity of a physical quantity in
\qcpt~is different from that in standard $\chi$PT, one
will expect its extrapolation to chiral limit in quenched lattice
calculations to be unreliable.
{}From a more general point of view, standard
 and quenched chiral perturbation
theories can be regarded as two extremes of a (discrete) spectrum of
theories with different degrees of quenching.
This connection is made possible
in partially quenched chiral perturbation
theory (PQ$\chi$PT) \cite{BG2}, which is the low energy effective theory
of QCD with $n$ quarks and $k$ ghost ($n \ge k$)~\footnote{
Various aspects of partially quenched QCD has been studied recently
in~\cite{pentronzio, sharpe}.}.
Standard and
quenched chiral perturbation theories are recovered by tuning
$k = 0$ and $n$ respectively.

Since its invention, \qcpt~has been used to study various hadrons
such as
Goldstone bosons \cite{S1, S2, BG1}, baryons \cite{sharpebaryon},
heavy mesons \cite{sharpeheavy} and exactly soluble two-dimensional
QED~\cite{eichten}.
Recently, Booth {\it et.~al.}~\cite{F}~has formulated \qcpt~for vector
mesons.
Here we will extend these works in two directions.
We will formulate PQ$\chi$PT for vector mesons and study their
non-analytic singularities in chiral loop corrections.
We will also show that, in the heavy mass expansion formalism, it is
straightforward to generalize the chiral perturbation theory to
$2^{++}$ tensor mesons or mesons with even higher spins.
The relationship of the present work to previous literature is
summarized in the following table:

\bigskip

\centerline{\vbox{
\halign{\hfil#\qquad&\hfil#\hfil&\qquad\hfil#\hfil&\qquad\hfil#\hfil\cr
$J^{PC}$&$\chi$PT&PQ$\chi$PT&Q$\chi$PT\cr
$0^{-+}$&Gasser, Leutwyler\cite{GL}&Bernard, Golterman\cite{BG2}&
Sharpe\cite{S1,S2}\cr
$1^{--}$&Jenkins, Manohar, Wise\cite{JMW}&Chow, Rey(this paper)&
Booth, Chiladze, Falk\cite{F}\cr
$2^{++}$&Chow, Rey\cite{CR}&Chow, Rey(this paper)&Chow, Rey(this
paper)\cr}}}

\bigskip

This paper is organized as follows.
In the next section, we will formulate PQ$\chi$PT of vector mesons.
We will then illustrate its application to calculate the chiral one-loop
correction to $\rho$-meson mass
in Section 3, and compare it with the results
from Q$\chi$PT and unquenched $\chi$PT.
Lastly, generalization to tensor mesons will be treated in Section 4.

\section{Partially Quenched QCD for Vector Mesons}
\setcounter{equation}{0}
In this section, we review partially quenched QCD~\cite{BG2} and
construct
chiral perturbation theory for vector mesons.

Partially quenched QCD consists of $n$ quarks $q_i$ and $k$ ghost-quarks
$\tilde q_j$. Quark masses $m_i (i = 1, \cdots, n)$ are completely
arbitrary and the ghost-quark masses are fixed to be equal to the masses
of the last $k$ quarks, viz. ${\tilde m}_j = m_{n-k + j}, (j = 1,
\cdots k)$. As such, partially quenched QCD contains the first $(n-k)$
unquenched quarks and the remainder $k$ quenched ones.
The full graded chiral
symmetry of the partially quenched QCD is the semi-direct product
[ $\sunk_L \times \sunk_R ] \times \u1$. The additional
axial $\u1$ is broken by the QCD anomaly (for $N_c < \infty$).
In the notation of~\cite{BG2} we will refer this theory as 
\sunk-theory. Note that \sunk-theory is expected to interpolate between 
fully unquenched QCD described by \sun0-theory and fully quenched QCD 
described by \sunn-theory.

\subsection{Goldstone Meson Multiplet Sector}

PQ$\chi$PT for Goldstone mesons was first studied in Ref.~\cite{BG2}.
The summary below is just a brief sketch of their work and the reader is
encouraged to the original paper for details.

Goldstone meson fields can be written as a (n+k)$\times$(n+k)
unitary matrix field $\Sigma$ defined as:
\be
\Sigma \equiv \exp ( 2 i \Phi / f), \hskip1cm
\Phi \equiv \left( \begin{array}{cc} \phi & \chi^\dagger \\
                   \chi & {\tilde \phi} \end{array} \right).
\ee
In terms of quarks and ghost-quarks, $\phi \approx (q_i {\overline
q}_j),
{\tilde \phi} \approx ({\tilde q}_i {\overline {\tilde q}}_j),
\chi \approx ({\tilde q}_i {\overline q}_j)$. Each of them are
(n$\times$n), (k$\times$k) and (k$\times$n) matrices respectively.
For $n = N_F = 3$, for example, $\phi$ is the Goldstone boson nonet
\be
\phi = \left( \begin{array}{ccc}
{\pi^0 \over \sqrt 2} + {\eta \over \sqrt 6} & \pi^+ & K^+ \\
\pi^- & -{\pi^0 \over \sqrt 2} + {\eta \over \sqrt 6} & K^0 \\
K^- & {\overline K}^0 & - {2 \eta \over \sqrt 6} \\
\end{array} \right)
+ {{\bf I} \over \sqrt 3} \eta'.
\ee
Under the chiral $\sunk_L \times \sunk_R$,
\be
\Sigma \rightarrow L \, \Sigma \, R^\dagger
\ee
where $L \in \sunk_L, \,\, R \in \sunk_R$, and under charge conjugation
$C$,
\be
C \, \Sigma \, C^{-1} = + \Sigma^\dagger.
\ee
The (n+k)$\times$(n+k) quark mass matrix is given by
\be
{\cal M}_{ij} = {\rm diag} (m_1, \cdots, m_k, m_{k+1}, \cdots,
m_n; m_{k+1} \cdots, m_n),
\ee
viz. the ghost quarks are degenerate in mass with $k$-flavors of
the quarks. We also introduce
\be
{\cal M}_\xi \equiv {1 \over 2} (\xi^\dagger {\cal M} \xi^\dagger
+ \xi {\cal M} \xi).
\ee

Interactions among Goldstone boson fields are described by the
chiral Lagrangian:
\bee
{\cal L}
= &&
F_8 ({\rm Str} \ln \Sigma) {\rm Str} (\partial_\mu \Sigma \partial^\mu
\Sigma^\dagger) + F_0 ({\rm Str} \ln \Sigma) {\rm Str} (\partial_\mu
\ln \Sigma) {\rm Str} (\partial^\mu \ln \Sigma^\dagger)
\nonumber \\
&&\qquad+ V_8 ({\rm Str} \ln \Sigma) {\rm Str} ({\cal M}_\xi)
 + V_0 ({\rm Str} \ln \Sigma).
\label{x}
\eee
Up to quadratic order in interactions~\cite{BG2} ,
\bee
F_8 ({\rm Str} \ln \Sigma) &=&  {f^2 \over 8} + \cdots
\nonumber \\
F_0 ({\rm Str} \ln \Sigma) &=&  {A_0 \over 6} + \cdots
\nonumber \\
V_8 ({\rm Str} \ln \Sigma) &=& { m_\pi^2 f^2 \over 4 m_q} + \cdots
\nonumber \\
V_0 ({\rm Str} \ln \Sigma) &=&  {\mu^2_0 f^2 \over 24 }
({\rm Str} \ln \Sigma) ({\rm Str} \ln \Sigma)^\dagger
+ \cdots.
\label{quadratic}
\eee
The main difference between quenched and unquenched QCD is the presence
of $F_0, V_0$ interactions that depend on
${\rm Str} \ln \Sigma$. In unquenched QCD,
this field corresponds to heavy $\eta'$-meson and decouples from
low-energy dynamics. In quenched QCD, since the $\eta'$-meson remains
light, their interactions has to be taken into account.

From the Lagrangian Eq.(\ref{x}), one can obtain propagators of the
Goldstone meson multiplets.
Flavor-charged Goldstone mesons have the same kinetic terms, hence,
propagator structures as those of full QCD.
On the other hand, for the flavor-neutral Goldstone mesons,
non-decoupling of ${\rm Str} \ln \Sigma$ field gives rise to
non-standard form of the kinetic term.
For simplicity, we will
consider the ``degenerate \sunk-theory'', where all
the $m_i (i = 1, \cdots, n)$ masses are equal.
In Euclidean momentum space, for $\alpha = 0$, the propagators of the
flavor-neutral Goldstone bosons are given by
\be
[G^{-1}]_{ij} = \delta_{ij} \, (p^2 + m_\pi^2) \epsilon_i
+ {\mu^2_0 \over 3} \epsilon_i \epsilon_j.
\ee
(Non-zero $A_0$ can be easily reinstated by shifting $\mu^2_0
\to \mu^2_0
+ A_0 \,  p^2$.)
The grading index $\epsilon_i$ is such that
$\epsilon (q_i) = + 1, \epsilon ({\tilde q}_j) = -1$.
The corresponding propagator takes an extremely simple form
\be
G_{ij} = \Big[ {\delta_{ij} \epsilon_i - 1/(n-k) \over p^2 + m_\pi^2}
+ {1/(n-k) \over p^2 + m_\pi^2 + (n-k)\, \mu^2_0/3}
\Big].
\label{pro}
\ee
The propagator is a sum of two simple-pole contributions:
one with equal mass to all meson multiplets and one with a shifted
mass including the singlet contribution.
Also note that, as $n \rightarrow k$, the shifted pole moves back to the
pion pole.
The propagator will then have a double pole at $m_\pi^2$, which is a
well-known result in \qcpt~\cite{BG1}.

\subsection{Vector Meson Multiplet Sector}

Standard chiral perturbation theory for vector mesons has been
formulated by Jenkins, Manohar and Wise in Ref.~\cite{JMW} and the
quenched counterpart by Booth, Chiladze and Falk in Ref.~\cite{F}.
Here we will construct the partially quenched theory.

Vector meson multiplet is described by (n+k)$\times$(n+k) graded matrix
field, much the same way in structure as the Goldstone meson multiplet:
\be
{\cal N}_\mu = \left( \begin{array}{cc} {\cal V} & \psi \\
\psi^\dagger & {\tilde {\cal V}} \end{array} \right)_\mu
\ee
where ${\cal V}_\mu$ is the usual vector meson matrix field, which
for $n=N_F = 3$ is given by
\be
{\cal V}_\mu = \left( \begin{array}{ccc}
{\rho^0 \over \sqrt 2} + {\phi^8 \over \sqrt 6} & \rho^+ & K^{*+} \\
\rho^- & - {\rho^0 \over \sqrt 2} + {\phi^8 \over \sqrt 6} & K^{*0} \\
K^{*-} & {\overline K}^{*0} & - { 2 \phi^8 \over \sqrt 6} \end{array}
\right)_\mu + {{\bf I} \over \sqrt 3} S_\mu
\ee
Under the $ [\sunk_L \times \sunk_R ] \otimes \u1$ graded chiral
symmetry,
\be
{\cal N}_\mu \rightarrow U \, {\cal N}_\mu \, U^\dagger
\ee
and under charge conjugation,
\be
C {\cal N}_\mu C^{-1} = - {\cal N}^T_\mu.
\ee

We treat the vector meson multiplet as heavy, static source, which
was previously utilized for conventional $\chi$PT for vector
mesons~\cite{JMW} and for Q$\chi$PT for vector mesons~\cite{F}.
In this formalism, the static vector
meson propagates with a fixed four-velocity $v_\mu$, $v^2 = 1$ and
interacts with soft Goldstone multiplets. Three polarization states
of vector mesons are perpendicular to the propagation direction,
viz. $v \cdot {\cal N} = 0$. The chiral Lagrangian which describes
the interactions of the vector meson multiplet with the soft Goldstone
meson multiplet consists of three parts. At leading order in derivative
and quark mass perturbations, they are
\be
{\cal L}_V = {\cal L}_{\rm kin} + {\cal L}_{\rm int} + {\cal L}_{\rm
mass}
\ee
where
\bee
{\cal L}_{\rm kin} &=& - i {\rm Str} ( {\cal N}^\dagger_\mu
v \cdot {\cal D} {\cal N}_\mu )
- i A_1 ({\rm Str} {\cal N}_\mu^\dagger )  v \cdot {\cal D} ({\rm Str}
{\cal N}_\mu)
\label{kinterm} \\
{\cal L}_{\rm mass}
&=& {\overline \mu} \,  {\rm Str} ({\cal N}^\dagger_\mu {\cal N}_\mu)
+ \mu_1 \,  ({\rm Str} {\cal N}^\dagger_\mu) ({\rm Str} {\cal N}_\mu)
\label{massterm} \\
&+& \lambda_1 \Big( ({\rm Str} {\cal N}_\mu^\dagger) ({\rm Str}
{\cal N}_\mu {\cal M}_\xi) + {\rm h.c.} \Big)
+ \lambda_2 {\rm Str} (\{ {\cal N}_\mu^\dagger, {\cal N}_\mu \}
{\cal M}_\xi ) \, .
\nonumber \\
\eee
Here, covariant derivative is defined by
\be
{\cal D}_\mu = \partial_\mu + [V^\mu, \,\,\,\,\,]
\ee
and
\be
V^\mu = {1 \over 2} (\xi \partial^\mu \xi^\dagger +
\xi^\dagger \partial^\mu \xi ), \hskip1cm
A^\mu =  {1 \over 2} (\xi \partial^\mu \xi^\dagger -
\xi^\dagger \partial^\mu \xi )  .
\ee
The second terms in Eqs.~(\ref{kinterm}, \ref{massterm} ) are new
interactions present in (partially) quenched QCD. The first term in
Eq.~(\ref{massterm})
corresponds `residual mass' ${\overline \mu}$
of vector meson multiplets. By a suitable
reparametrization transformation~\cite{reparametrization},
it is always possible to remove the
residual mass. Using this freedom, we will choose ${\overline \mu} = 0$
throughout this paper.
The last two terms in Eq.~(\ref{massterm}) correspond to SU(3) isospin
breaking due to quark masses.
We will not use these terms in this paper.

The propagator of heavy vector mesons is similar to Goldstone meson
multiplets.
For flavor non-diagonal vector mesons, the propagator is
\be
G_{\mu \nu} (k) = \Pi_{\mu \nu} {1 \over v \cdot k}
\ee
where $k^\mu$ denotes residual momentum of vector meson and
\be
\Pi_{\mu \nu} \equiv (v^\mu v^\nu - g^{\mu \nu})
\ee
is the projection operator. For flavor-diagonal vector meson
multiplets, the propagator is
\be
G_{ij, \mu \nu} (k) = \Pi_{\mu \nu}
\Big[ {\delta_{ij} \epsilon_i - 1/(n-k) \over v \cdot k}
 + {1 / (n-k) \over v \cdot k + (n-k) \mu_1 } \Big],
\label{xxx}
\ee
in which we have set $A_1 = 0$. Non-zero $A_1$ can be incorporated
by shifting $\mu_1 \rightarrow \mu_1 + A_1 v \cdot k$.

There are four-types of chiral invariant interactions between
vector meson multiplets and Goldstone boson multiplets that are
consistent with graded chiral symmetry:
\bee
{\cal L}_{\rm int} &=& i g_1 ({\rm Str} {\cal N}^\dagger_\mu)
({\rm Str} {\cal N}_\nu A_\lambda ) v_\sigma \epsilon^{\mu \nu \lambda
\sigma} + {\rm h.c.}
\nonumber \\
&& + i g_2 {\rm Str} (\{ {\cal N}^\dagger_\mu, {\cal N}_\nu \}
A_\lambda)
v_\sigma \epsilon^{\mu \nu \lambda \sigma}
\nonumber \\
&& + i g_3 ({\rm Str} {\cal N}^\dagger_\mu) ({\rm Str} {\cal N}_\nu )
({\rm Str} A_\lambda) v_\sigma \epsilon^{\mu \nu \lambda \sigma}
\nonumber \\
&& + i g_4 {\rm Str} ({\cal N}^\dagger_\mu {\cal N}_\nu )
({\rm Str} A_\lambda) v_\sigma \epsilon^{\mu \nu \lambda \sigma}.
\label{int}
\eee

\section{Non-analytic Chiral Correction to Vector Meson Mass}

To illustrate the application of PQ$\chi$PT of vector mesons, we will
calculate the mass correction of the $\rho$ meson in PQ$\chi$PT.
The $\rho$ meson mass is highly important in lattice calculation as
it is often used to set the overall mass scale.
The non-analytic correction to $m_\rho$ in Q$\chi$PT has been calculated
in Ref.~\cite{F}, where it is shown to be very different from its
counterpart in the standard $\chi$PT \cite{JMW,F}.
We will see below that in a sense the PQ$\chi$PT result is intermediate
between these two extreme cases.

\subsection{Remarks on Parameters of PQ$\chi$PT Lagrangian}
The chiral one-loop corrections to $\rho$-meson mass
 come from the diagrams shown in Fig.~1.
Fig.~1a is from the expansion of ${\cal L}_{\rm kin}$.
The resulting one-loop tadpole also contains both regular and hairpin
insertion contributions. However, because
the interaction vertex is proportional to the vector meson four-velocity
$v^\mu$, this partially quenched one-loop tadpole contributes only to
the wave function renormalization but not to the vector meson mass
correction\footnote{This chiral one-loop contribution was not considered
in
\cite{F}. }.
Such tadpole diagrams also comes from the quark mass matrix terms in the
Lagrangian and in general may lead to non-trivial mass correction.
However, we will be working in the degenerate mass limit, in which these
tadpole diagrams do not contribute.
Fig.~1b comes from the interaction Lagrangian Eq.(\ref{int})
and depends on
the values of $g_{1,2,3,4}$, as well as the hairpin parameters $A_0$,
$\mu_0^2$ in the Goldstone meson sector and $A_1$, $\mu_1$ in the vector
meson sector.
Just as the coupling constants in normal $\chi$PT, they are undetermined
parameters in the theory.
To make the matter worse, the values of these parameters of PQ$\chi$PT
need not to be the same as those in QCD, and hence cannot be extracted
from experimental data. In principle, one may be
able to extract the value of these parameters from lattice
simulation data, which may accumulate enough in the foreseeable future.
Evidently it is difficult to extract any useful information from
the theory with so many undetermined parameters. Because of this 
complication, we pick particular values for these parameters partly motivated
by comparison with standard QCD chiral Lagrangian in the large $N_c$ limit 
and hope that the set of values we picked is
``generic'', viz. no accidental cancelation or enhancement takes place.
In order to make our presentation as clear as possible, we restrict our
investigation to the above truncated set of interactions and relegate
a complete and detailed study retaining all the relevant couplings to a 
separate paper.

Since we are going to compare our results to the Q$\chi$PT counterparts,
we will choose the same hairpin parameters as adopted in
Ref.~\cite{F}, i.e.,
$A_1 = 0$, $\mu_1=0$, while $\mu_0^2/3 = (400 {\rm MeV})^2$ and $A_0/3 = 0.2$.
Note that these choices are fundamentally arbitrary; there is no reason
that the parameters have to have identical values in these two different
theories, although this seems to be the most ``natural'' and
``unbiased''
choice and make comparison easy.
We will also follow Ref.~\cite{F} in setting $g_{3,4}=0$ and $g_2 =
0.75$,
the latter having the same value of the counterpart in normal $\chi$PT
\cite{JMW}.
We emphasize again that there is {\sl no} reason why we should set these
coupling parameters the same between the two $\chi$PT's. 
We have chosen to differ from Ref.~\cite{F} in setting $g_1=0$
instead
of 0.75.
This choice is based on the observation that generically there will be
interference terms proportional to $g_1g_2$, hence, that the result will
be highly sensitive to the {\sl relative sign} between $g_1$ and $g_2$,
which is again theoretically undetermined.
Since we have no way to discern whether the interference should be
constructive or destructive, we will
choose to set $g_1=0$ in order to obtain
{\sl interference-independent predictions}.
Again, we emphasize that this is an arbitrary choice of parameters.
If desired, one can easily perform an analogous calculation with a
different set of parameters.
Nevertheless, we expect that this choice retains essential physics and that
the results are at least qualitatively correct.

\begin{figure}
\vspace{1.5cm}
\hspace{5cm}
\epsfig{file=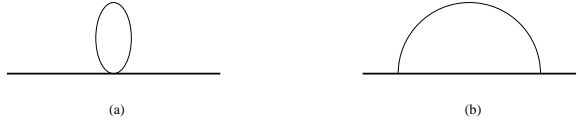, height=3in,angle=-90,clip=}
\vspace{0.5cm}
\caption{Chiral one-loop diagrams to vector meson two-point function. 
Horizontal lines denote vector mesons and upper internal lines are 
Goldstone meson multiplets.}
\vspace{1.5 cm}
\end{figure}
\subsection{Chiral One-Loop Calculation}
Now we are at the stage for the calculation of the chiral one-loop
correction to the $\rho$ meson mass in PQ$\chi$PT.
All the diagrams involved have the same one-loop Feynman integral
structure\footnote{Note that the same object is denoted as ${\cal
I}_1(m)$ in Ref.~\cite{F}.} :
\begin{equation}
{\cal I}_1 (m^2) = - {1 \over 12 \pi f^2} m^3.
\end{equation}
For concreteness
we will calculate the mass correction of a charged $\rho$
meson.
(The calculation would be somewhat more complicated with a neutral
$\rho$ meson,
but because of the isospin symmetry, the mass correction of all
$\rho$ mesons should be the same.)
With the choice that only $g_2\neq 0$, the ghost-antighost meson does
not
contribute, and there are four different types of contribution to the
mass correction $\Delta m_\rho$:

\bigskip

\noindent
(1) \tt Intermediate Goldstone meson is
quark-antighost or ghost-antiquark
meson. \rm

The contribution is given by
\be
\Delta m_\rho = - g^2_2 \, \cdot  \, 2 k \, \cdot \, {\cal I}_1
(m_\pi^2),
\ee
where the factor $2k$ follows from the contraction of flavor indices.

\medskip

\noindent
(2) \tt Intermediate Goldstone meson is flavor-charged \rm
(for $n = 2$, $u \bar d=\pi^+$ and $d \bar u=\pi^-$).

The contribution is given by
\be
\Delta m_\rho = + g_2^2 \, \cdot \, 2 (n-1) \, \cdot \,
{\cal I}_1 (m_\pi^2).
\ee

\medskip

\noindent
(3) \tt Intermediate Goldstone meson is flavor-neutral,
{\rm (for $n = 2$, $u \bar u$ and $d \bar d$, which are linear
combinations
of $\pi^0$ and $\eta'$)}, and the propagator is given by the first term
in
Eq.~(\ref{pro}).  \rm

The contribution is given by
\be
 \Delta m_\rho = + g_2^2 \, \cdot \, 2 \, \cdot \,
\Big(1 - {2 \over \Delta n} \Big) \, \cdot \, {\cal I}_1 (m_\pi^2).
\ee
where $\Delta n \equiv ( n - k )$.

\medskip

\noindent
(4) \tt Intermediate Goldstone meson is flavor-neutral, and the
propagator
is given by the second term in Eq.~(\ref{pro}). \rm

For this term the pole in the Goldstone boson propagator is shifted,
and the contribution is given by
\be
\Delta m_\rho = + g_2^2 \, \cdot \, 2 \, \cdot \,
\Big( {2 \over \Delta n} \Big) \cdot \Big( {1\over 1 + \Delta n \cdot
A_0 /3} \Big) \, \cdot \,
{\cal I}_1 \Big({m_\pi^2 + \Delta n \cdot \mu^2_0/3 \over
1 + \Delta n \cdot A_0 /3} \Big).
\ee

\bigskip

The total chiral one-loop correction is a sum of these four
contributions:
\be
\Delta m_\rho = 2 g_2^2 \, \Big[
\Delta n \, {\cal I}_1(m_\pi^2)
- {2\over \Delta n} \, {\cal I}_1(m_\pi^2)
+ {2\over \Delta n} \Big( {1 \over 1+\Delta n \cdot A_0 /3} \Big) \,
{\cal I}_1 \Big(
{m_\pi^2 + \Delta n \cdot \mu^2_0 /3 \over 1 + \Delta n \cdot
A_0 /3} \Big) \Big].
\label{result}
\ee
Note that the final result depends only on $\Delta n$, which counts the
difference in the number of quarks and the number of ghosts, but not
$n$ or $k$ seperately.
Physically this reflects an obvious fact that physical
quantities should be unchanged upon introduction of a degenerate set of
extra quark and extra ghost,
as their contribution should cancel out completely.

Finally, we note that our result can be expressed in terms of just
${\cal I}_1$, which also appears in the standard $\chi$PT result.
(See below for a comparison between PQ$\chi$PT and $\chi$PT results.)
The ``new chiral singularities'' or ``quenched infared divergences''
which
appear in Q$\chi$PT and are denoted by ${\cal I}_{2,3,4}$ in
Ref.~\cite{F},
do not appear.
This is in full agreement with the Bernard--Golterman's third
theorem~\cite{BG2},
which states that {\sl quenched infared divergences appear if and only
if
one or more of the valence quarks are fully quenched.}
Since the theory we are considering is only partially quenched, we do
not
see these new chiral singularities.

\subsection{$\Delta n = 0$ Case -- Fully Quenched QCD Limit}
As mentioned above, partially quenched QCD
 was introduced to bridge between the two
extreme cases, namely fully quenched and unquenched QCD theories.
One would expect, by setting $\Delta n = 0$, the Q$\chi$PT results
should be recovered.
In particular, the non-analytic
quenched infared singularities should reappear.
Can we see this explicitly from our results?

The answer is a resounding yes.
Let us see how this arises.
Note that Eq.~(\ref{result}) can be re-expressed as
\begin{equation}
\Delta m_\rho = 2g_2^2 \, \Big[\,
\Delta n \cdot {\cal I}_1(m_\pi^2) + {2\over \Delta n} 
\cdot \Big( \, \Big({1\over 1+\Delta n A_0/3} \Big) \,
{\cal I}_1 \Big({m_\pi^2 + \Delta n \cdot \mu^2_0/3
\over 1 + \Delta n \cdot A_0 /3} \Big) -
{\cal I}_1(m_\pi^2) \Big) \Big].
\end{equation}
When $\Delta n \to 0$, the first term vanishes while the second term
becomes
a derivative;
\begin{equation}
\Delta m_\rho = 4 g_2^2 \, {d\over d \Delta n} \,
\Big[ \, \, {1\over 1+\Delta n \cdot A_0 /3} \cdot
{\cal I}_1 \Big( \,
{m_\pi^2 + \Delta n \cdot \mu^2_0 /3 \over 1 + \Delta n \cdot A_0 /3}
\, \Big)\,\, \Big]_
{\Delta n = 0}.
\end{equation}
Now it is clear how the Bernard--Golterman's third theorem breaks down
(as predicted by Bernard and Golterman) in the fully quenched limit.
While the contributions from both the pion pole and the shifted pole are
of the functional
form ${\cal I}_1$, in the fully quenched limit the shifted pole
returns back
to its unshifted position and produces a derivative term which is
{\it not\/} of the form ${\cal I}_1$.
Expanding the derivative explicitly, one finds that
\begin{eqnarray}
{\cal I}_2(m_\pi^2) &\equiv & {d\over d \Delta n} \,
\Big[ \, {1\over 1+\Delta n \cdot A_0 /3} \cdot
{\cal I}_1\Big(
{m_\pi^2 + \Delta n \cdot \mu^2_0 /3 \over 1 + \Delta n \cdot
A_0 /3})\Big]_
{\Delta n = 0}
\nonumber \\
&= & \, {1\over 12\pi f^2} \, \Big({3\over2} \cdot {\mu^2_0 \over3}
m_\pi
- {5\over2}\cdot {A_0 \over3} \cdot m_\pi^3 \Big),
\end{eqnarray}
and the term linear in $m_\pi$ is the anticipated non-analytic
quenched infared singularity.
The mass correction in the fully quenched limit is
\begin{equation}
\Delta m_\rho = 4 \, g_2^2 \, {\cal I}_2(m_\pi^2),
\end{equation}
which agrees perfectly
with Eq.~(3.23) of Ref.~\cite{F}\footnote{Note that  $M_0^2$
and $A_0$ in Ref.~\cite{F} are our $\mu^2_0/3$ and $A_0 /3$
respectively.
Also recall
 that we have kept $g_2$ as the only non-vanishing coupling.}.
We thus have established that PQ$\chi$PT does reproduce Q$\chi$PT
results in the
fully quenched limit $\Delta n = 0$.
\subsection{$k=0$ Case -- Unquenched Limit}

Let's now turn our attention to the other end of the limits.
Does it reproduce the standard
 $\chi$PT results in the unquenched limit $k=0$?

This time the answer is clearly no!
In normal $n = 2$ $\chi$PT, we have only pion loops and hence do not
expect a contribution from a shifted pole.
In fact, the $\chi$PT result is straightforwardly calculated to be
\begin{equation}
\Delta m_\rho = g_2^2 \, \Big(2n - {4\over n} \Big) \,
{\cal I}_1(m_\pi^2)
\label{ChPT}
\end{equation}
which, as expected, does not contain
any contribution from a shifted pole.
So the question is, why have we failed to reproduce the unquenched
QCD results?

This puzzle will be resolved by noting how we have obtained
the $\chi$PT result quoted above.
Let's imagine a world in which $\eta'$ is degenerate with all the
other Goldstone bosons (for example, $N_c \rightarrow \infty$ world).
The mass correction comes from Feynman diagram in Fig. 1b,
and each diagram
gives rise to a mass correction $g_2^2 \cdot {\cal I}_1(m_\pi^2)$.
In each diagram we have an internal quark loop, which can take $n$
flavors and 2 orientations (clockwise/anticlockwise), so there are
in total $2n$ diagrams.
So naively one would expect $\Delta m_\rho = g_2^2 \cdot 2n \cdot
{\cal I}_1(m_\pi^2)$.
But this is not what we get from $\chi$PT, as this naive result includes
contribution from $\eta'$ loops, which has been integrated out in the
standard $\chi$PT \cite{JMW}.
It is not difficult to see that the $\rho\rho\eta'$ coupling is
$2g_2/\sqrt{n}$, so the $\eta'$ contribution is $g_2^2 (4/n)
{\cal I}_1(m_\pi^2)$.
Result (\ref{ChPT}) is obtained exactly
when the $\eta'$ contribution is subtracted
from the naive result.

Indeed, Eq.~(\ref{ChPT}) agrees with the $n=2$ result \cite{F},
\begin{equation}
\Delta m_\rho = g_2^2 \cdot 2 \cdot {\cal I}_1(m_\pi^2)
\end{equation}
and the $n=3$ result \cite{F,JMW}
\begin{equation}
\Delta m_\rho = g_2^2 \big( \, 2 \, {\cal I}_1(m_\pi^2)
+ 2 \, {\cal I}_1(m_K^2)
+ {2\over3} \, {\cal I}_1(m_\eta^2)\big) = g_2^2 \cdot {14\over3}
\cdot {\cal I}_1 (m_\pi^2) \qquad \hbox{when $m_\pi=m_K=m_\eta$.}
\end{equation}
It also reproduces exactly the ${\cal I}_1(m_\pi^2)$ term in
Eq.~(\ref{result}).
Now it becomes clear what the remaining contribution
from the shifted pole means physically:
it is the reinstatement of $\eta'$ contribution, and the shifting of the
pole just reflects that, in the real world, the $\eta'$ mass is shifted
with respect to the pion mass due to the necklace diagrams.
In other words, we can rearrange Eq.~(\ref{result}) in the following
way,
\begin{equation}
\Delta m_\rho = g_2^2 \,\Big( 2n - {4\over n} \Big) \,
{\cal I}_1(m_\pi^2)
+ g_{\rho\rho\eta'}^2 \, {\cal I}_1(m_{\eta'}^2),
\end{equation}
where
\begin{equation}
g_{\rho\rho\eta'}^2 = {4g_2^2/ Zn}, \qquad
m_{\eta'}^2 = (m_\pi^2 + n \mu^2/3)/Z, \qquad
{\rm where} \qquad Z = (1 + n \, A_0/3),
\end{equation}
which are the coupling, mass and wavefunction renormalization of the
$\eta'$ meson. Being organized this way, the first term of Eq.(38) exhibits
exactly of the same functional form as the standard $\chi$PT result. Note
that
numerical value of respective coupling parameter $g_2$ in PQ$\chi$PT 
and in standard $\chi$PT differ generically each other. If the $\eta'$ meson
were integrated out instead of being truncated, its effect will be reflected
to finite additive renormalization of the coupling parameters so that
their values for PQ$\chi$PT becomes numerically identical to those of $\chi$PT.

It is of interest to compare the above result with the Bernard-Golterman's
first theorem. According to the theorem, 
{\sl in the subsector where all valence
quarks are unquenched, the \sunk~theory is completely equivalent to a
normal, completely unquenched \su-0~theory.}
We found that this is indeed true except that, compared to standard $\chi$PT,
this ``completely unquenched \su-0~theory'' retains an $\eta'$
meson which may (and does) contribute to chiral loop corrections.
In fact, Bernard and Golterman~\cite{BG2} have already noted this aspect
from their study of chiral perturbation theory for pion self-energy
correction. 
In a low-energy effective theory, when treating not-so-heavy field 
excitations, one may either retain them or integrate out. As such, it 
should be viewed as a matter of choice whether one includes the $\eta'$ 
in the effective theory.  As with Bernard and Golterman, we have noted 
that effective theory of unquenched \su-0~theory is the one which 
retains $\eta'$ explicitly.  It should then be
straightforward to identify Goldstone meson and $\eta'$ contributions
separately in a physical quantity, as is exemplified from the above 
calculation. 
 
\subsection{Numerical Comparison}
To conclude this section, we provide plots of $\Delta m_\rho$ as a 
function of pion mass in the following theories in Fig.~2. Recall that
we have chosen the quenching parameters as $A_0/3 = 0.2$ and $\mu_0^2/3 = 
(400 \, MeV)^2$.

0)  Standard $n=N_F = 2$ $\chi$PT,

1)  PQ$\chi$PT with $(n,k)=(2,0)$, which is identical with $\chi$PT
with $\eta'$ contribution,

2)  PQ$\chi$PT with $(n,k)=(2,1)$,

3)  PQ$\chi$PT with $(n,k)=(2,2)$, which is identical with Q$\chi$PT.

\begin{figure}[t]
\vspace{-9cm}
\hspace{0.cm}
\leftline{\vspace{0cm} \hspace{1.5cm} \epsffile{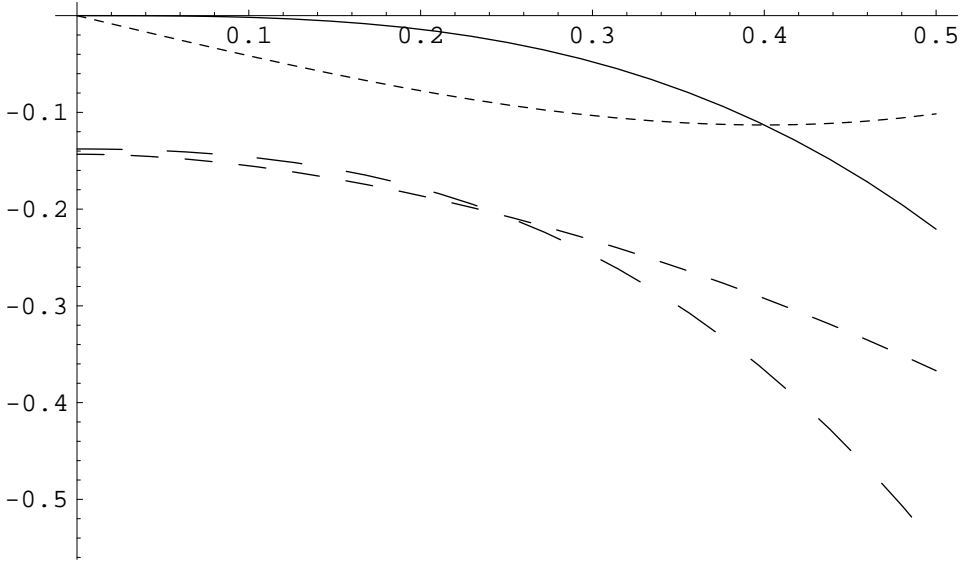}}
\vspace*{15.5cm}
\caption{Chiral one-loop correction $\Delta m_\rho$ (in unit of GeV) as a function of
pion mass in the range (0, 0.5) GeV. 
Solid line is the standard QCD, small dashed line is the (2,2) theory, 
medium dashed line is the (2,1) theory and large-dashed
line is the (2,0) theory.}
\end{figure}
\vspace{0cm}
\begin{figure}[t]
\vspace{-9cm}
\hspace{0cm}
\leftline{\hspace{1.5cm} \epsffile{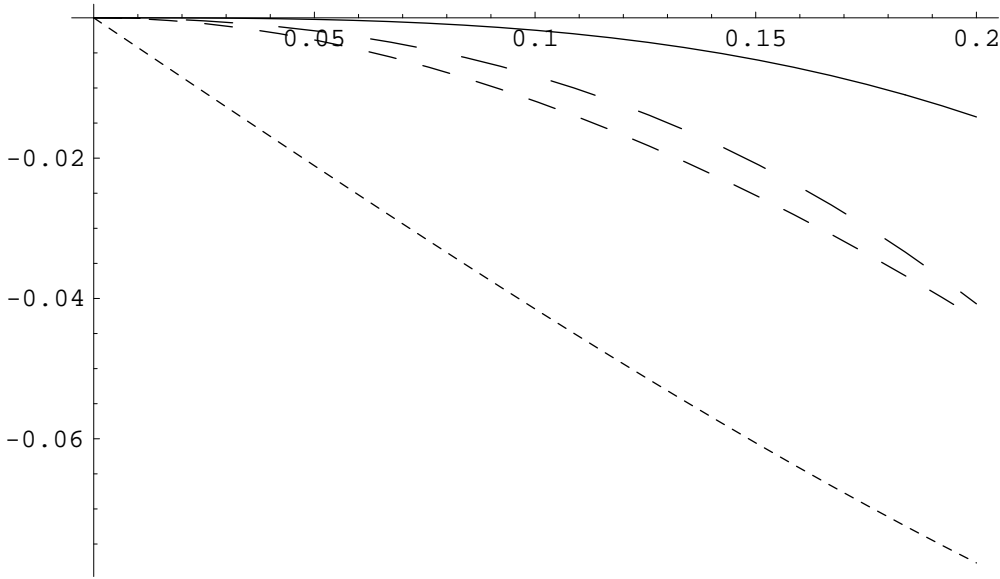}}
\vspace*{15.5cm}
\caption{$(\Delta m_\rho - \Delta m_\rho|_{m_\pi = 0})$ in unit of GeV 
as a function of pion mass in the range (0, 0.2) GeV. 
Solid line is the standard QCD, small dashed line is the (2,2) theory, 
medium dashed line is the (2,1) theory and large-dashed
line is the (2,0) theory.}
\end{figure}
\vspace{0cm}

From the plot it is clear that PQ$\chi$PT is qualitative different from 
both $\chi$PT and Q$\chi$PT.  
In both $\chi$PT and Q$\chi$PT $\Delta m_\rho$ vanishes in the chiral 
limit, while it is not the case for PQ$\chi$PT.  
In fact, from Eq.~(\ref{result}) in the chiral limit, we find  
\begin{equation}
\Delta m_\rho\bigg|_{m_\pi^2=0} = {4g_2^2\over \Delta n} \cdot
\Big( {1\over 1+\Delta n \cdot 
A_0/3} \Big) \cdot 
{\cal I}_1\bigg({\Delta n \cdot \mu^2_0 /3\over1+\Delta n \cdot 
A_0/3}\bigg).  
\end{equation}
As mentioned in the previous sections, this contribution 
arises from heavy $\eta'$ 
loops, which could have been
 integrated out of the low energy effective theory to 
recapture the correct chiral limit of $\Delta m_\rho$.  
By integrating out the $\eta'$ meson in PQ$\chi$PT, parameters in the 
Lagrangian are renormalized, the details of which are beyond 
the scope of the present study.  In order to probe $\eta'$ meson chiral
loop correction and its pion mass dependence, we have subtracted 
$\Delta m_\rho|_{m_\pi^2=0}$ from Eq.(30). The corrected plots are shown in 
Fig.~3 {\sl assuming} tacitly that respective couplings are all the same. 
It is clearly verified, in the chiral limit, that PQ$\chi$PT does behave as a
continuous intrapolation between $\chi$PT and Q$\chi$PT as expected. It should
be emphasized again that, should we subtract the $\eta'$ contribution for a
given pion mass, the PQ$\chi$PT and $\chi$PT agree perfectly each other,
as discussed in detail in the previous subsection.
Thus, the difference between the solid and the large-dashed lines reflects
dependence of the chiral correction due to $\eta'$ meson to the pion mass.

\section{Partially Quenched QCD and Tensor Mesons}
\setcounter{equation}{0}
In Ref.~\cite{CR} $\chi$PT for tensor mesons have been developed.
It has been
 found that $\chi$PT for tensor mesons is very similar to the
vector meson counterpart, the only difference being the extra Lorentz
indices contracted trivially. This is the reflection of decoupling
from the dynamics of
spin and flavor quantum numbers
manifest in the heavy mass formalism that we have adopted for
describing these mesons.
This observation applies equally well to Q$\chi$PT and PQ$\chi$PT,
as the Lorentz
structure of these theories are identical with that in $\chi$PT.
For completeness, we will formally write down PQ$\chi$PT for tensor
mesons. The Q$\chi$PT can be recovered by setting $n=k$.

\subsection{PQ$\chi$PT for Tensor  Mesons}
Tensor meson multiplet is again described by (n+k)$\times$(n+k) graded
matrix field in
exactly the same manner as Goldstone meson and vector meson multiplets:
\be
{\cal N}_{\mu \nu} = \left( \begin{array}{cc}
{\cal V} & \psi \\ \psi^\dagger & {\tilde {\cal V}} \end{array}
\right)_{\mu \nu}.
\label{tensorfield}
\ee
For the multiplets, we use the same symbol as the vector meson
multiplet.
This should not cause any confusion: spin of the multiplet is easily
identified with the Lorentz indices.
In Eq.~(\ref{tensorfield}), ${\cal V}_{\mu \nu}$ denotes the usual
tensor meson field, which for $n=N_F = 3$ is given by
\be
{\cal V}_{\mu \nu} =  \left( \begin{array}{ccc}
{a^0_2 \over \sqrt{2} } + {f^{(8)}_2 \over \sqrt {6}} &
a_2^+ & K^{*+}_2 \\
a_2^- & - {a^0_2 \over \sqrt 2} + {f_2^{(0)} \over \sqrt 6} & K_2^{*0}
\\
K^{*2}_-  & {\overline K}^{*0}_2 & - {2 f^{(8)}_2 \over \sqrt 6}
\end{array} \right)_{\mu \nu} + {{\bf I} \over \sqrt 3} f_2^{(0)}.
\ee
Under the $[\sunk_L \times \sunk_R ] \otimes \u1$ graded chiral symmetry
\be
{\cal N}_{\mu \nu} \rightarrow U {\cal N}_{\mu \nu} U^\dagger
\ee
and under charge conjugation,
\be
C {\cal N}_{\mu \nu} C^{-1} = + {\cal N}^T_{\mu \nu}.
\ee

Following the construction developed in Ref.~\cite{CR}, we treat the
static tensor mesons as propagating with a fixed four-velocity $v_\mu,
\,\,\, v^2 = 1$ and as interacting with soft Goldstone meson multiplet
along the trajectory. By definition, these tensor mesons are symmetric
and traceless in Lorentz indices:
\be
{\cal N}_{\mu \nu} = {\cal N}_{\nu \mu},
\hskip1cm {\cal N}^\mu_\mu = 0.
\ee
Moreover, the polarizations of the tensor mesons are necessarily
orthogonal to the momentum, hence,
\be
v^\mu {\cal N}_{\mu \nu} = 0.
\ee

The chiral Lagrangian density which described the interactions of the
tensor meson multiplet with the low-momentum Goldstone meson multiplet
has the same structure as the vector meson multiplet case:
\be
{\cal L}_T = {\cal L}_{\rm kin} + {\cal L}_{\rm mass} +
{\cal L}_{\rm int} .
\ee
At leading order in the derivative and quark mass expansions and
in Euclidean space,
\bee
{\cal L}_{\rm kin} &=&
- {i \over 2} {\rm Str} ( {\cal N}^\dagger_{\mu \nu}
\, v \cdot {\cal D} \, {\cal N}_{\mu \nu} )
- {i \over 2} A_2 ({\rm Str} {\cal N}^\dagger_{\mu \nu} ) \,
v \cdot {\cal D} \, ({\rm Str} {\cal N}_{\mu \nu})
\nonumber \\
{\cal L}_{\rm mass} &=&
{{\overline \mu}_2\over2} \, {\rm Str} ({\cal N}^\dagger_{\mu \nu}
{\cal N}_{\mu \nu} ) +
{\tilde\mu_2 \over 2} ({\rm Str} {\cal N}^\dagger_{\mu \nu} )
({\rm Str} {\cal N}_{\mu \nu} )
\nonumber \\
&+&
{\tilde\lambda_1 \over 2}
\Big( ({\rm Str} {\cal N}^\dagger_{\mu \nu}) ({\rm Str}
{\cal N}_{\mu \nu} {\cal M}_\xi ) + {\rm h.c.} \Big)
+ {\tilde\lambda_2 \over 2}
{\rm Str} (\{ {\cal N}^\dagger_{\mu \nu}, {\cal N}_{\mu \nu}\}
{\cal M}_\xi ).
\eee
As in the vector meson multiplet, we turn off isospin breaking
quark mass perturbations and set the `residual mass' ${\overline \mu}_2
= 0$.

In terms of tensor projection operator
\be
\Pi^{\mu \nu, \alpha \beta}
\equiv (v^\mu v^\nu - g^{\mu \nu} ) ( v^\alpha v^\beta - g^{\alpha
\beta}) + {\rm permutations},
\ee
the tensor meson multiplet propagators are expressed as
\be
G_{\mu \nu, \alpha \beta} (k) = \Pi_{\mu \nu \alpha \beta}
{1 \over v \cdot k}
\ee
for flavor non-diagonal tensor meson multiplets, and
\be
G^{ij}_{\mu \nu, \alpha \beta} (k)
= \Pi_{\mu \nu \alpha \beta}
\Big[ {\delta_{ij} \epsilon_i - 1/(n-k) \over v \cdot k}
+ {1 / (n-k) \over v \cdot k + (n - k) \mu_2} \Big]
\ee
for flavor-diagonal multiplets. Again, in Eqs.~(\ref{pro}, \ref{xxx}),
we have set $A_2 = 0$. Non-zero $A_2$ is reinstated
by shifting $\mu_2 \rightarrow \mu_2 + A_2 \, v \cdot k$.

The chiral invariant interactions between tensor meson multiplets and
Goldstone meson multiplets is essentially the same form as those of
vector meson multiplets except contractions of Lorentz indices:
\bee
{\cal L}_{\rm int}
&=& i {\tilde g_1 \over 2} ({\rm Str} {\cal N}^\dagger_{\mu \alpha})
({\rm Str} {\cal N}_{\nu \alpha} A_\lambda ) v_\sigma
\epsilon^{\mu \nu \lambda \sigma} + {\rm h.c.}
\nonumber \\
&+& i {\tilde g_2 \over 2} {\rm Str} (\{ {\cal N}^\dagger_{\mu \alpha},
{\cal N}_{\nu \alpha} \} A_\lambda ) v_\sigma \epsilon^{\mu \nu \lambda
\sigma}
\nonumber \\
&+& i {\tilde g_3 \over 2} ({\rm Str} {\cal N}^\dagger_{\mu \alpha} )
({\rm Str} {\cal N}_{\nu \alpha}) ({\rm Str} A_\lambda) v_\sigma
\epsilon^{\mu \nu \lambda \sigma}
\nonumber \\
&+& i {\tilde g_4 \over 2} {\rm Str} ({\cal N}^\dagger_{\mu \alpha}
{\cal N}_{\mu \alpha} ) ({\rm Str} A_\lambda ) v_\sigma
\epsilon^{\mu \nu \lambda \sigma}.
\eee

Generalization to higher spin tensor meson multiplet is
completely straightforward and is left as an exercise to the reader.

\subsection{Chiral One-Loop Mass Correction to Tensor Mesons}
From the construction above, it should be
evident that the flavor structure decouples completely from the
Lorentz spin structure in both the vector and the tensor case.
Moreover, while tensor meson fields
carry one more Lorentz index than vector
meson fieldss,
the extra Lorentz indices are always contracted trivially in
the Lagrangian.
This has led us to an interesting observation in Ref.~\cite{CR}
that mass
corrections due to 1-loop effects in $\chi$PT for tensor mesons is
proportional to their counterparts for vector mesons, {\it up to
numerical equality of the respective coupling contants}.
Utilizing non-relativistic quark model, which is expected to be valid in
large $N_c$ limit, the proportionality constant has been calculated in 
Ref.~\cite{CR} to be 3/2. 
Since both the decoupling of flavor and Lorentz structures and the
trivial contraction of additional Lorentz indices also continues to
hold true for
(P)Q$\chi$PT, the proportionality result is also valid for (P)Q$\chi$PT.
We will state the result explicitly below:

\bigskip

{\bf Proposition:} {\sl In non-relativistic quark model, which is expected
to be a good approximation in large $N_c$ limit, chiral 1-loop corrections 
PQ$\chi$PT to tensor meson
masses are 3/2 times their counterparts in the vector meson masses,
with all the
coupling constants and hairpin parameters replaced by the corresponding
parameters in the tensor meson Lagrangian.  This result continues to
hold in both the fully quenched and the completely unquenched limit.}

\bigskip

If one assumes the parameters in the vector and tensor meson chiral
Lagrangians are the same, which we have argued to be a reasonable 
approximation in {\it non-relativistic quark model at large $N_c$ limit},
we deduce the following relation valid at chiral one-loop order
\begin{equation}
\Delta m_{a_2} = \textstyle{3\over2} \Delta m_\rho.
\label{relation}
\end{equation}
This observation entails an interesting consequence. In quenched
lattice QCD calculations,
when extracting light hadron spectrum, physical mass scale
is conveniently
normalized by identifying $\rho$-meson pole on the lattice with physical
$\rho$-meson mass. However, we have seen that vector meson mass is
plagued by quenched chiral logarithms in the chiral limit. As such,
it should be more desirable to use a combination of physical paramters
that is insensitive to the quenched approximation.
From the analysis above and Eq.~(\ref{relation}), we expect quenched
infared divergences to be
small for the combination $\Big( m_{a_2} - {3\over2} m_\rho \Big)$,
and one may consider using this combination to set the mass scale in
lattice
calculation\footnote{However, this will become practical only when the
statistical
uncertainties in the determination of tensor mesons is as small as
those of vector mesons.}.
In non-relativistic approach to lattice QCD~\cite{nrqcd}, 
the physical mass scale
is normalized by identifying S- and P-wave charmonium mass splitting
on the lattice with its Particle Data Group value. It has been noted
that this choice is quite insensitive to the quenching effect. We
suspect
that the underlying reason is similar to our proposal given above:
for heavy charmonium, spin and flavor decouples from the dynamics and
both S- and P-wave charmonium should receive similar chiral corrections.

\section{Discussions}
In this paper, we have developed quenched and partially quenched
chiral perturbation theory for vector and tensor mesons. 
We have formulated the quenched Lagrangian, and evaluated the chiral 
correction to $m_\rho$ under a specific choice of parameters that 
are partly motivated by large-$N_c$ limit.  
While the formal expression of $\Delta m_\rho$ may be modified if one 
also includes the effects of the hairpin parameters of the singlet 
vector meson and/or the $g_{1,3,4}$ couplings, we expect our analysis 
has captured the generic qualitative feature of PQ$\chi$PT.  
Just as the essential new physics (when compared to standard unquenched 
$\chi$PT) of Q$\chi$PT is the double pole in the singlet meson propagators, 
the essential new physics of PQ$\chi$PT is the shifted pole, which is 
carefully studied in this paper.  
Through this study we have clarified the connection between PQ$\chi$PT to its 
two extreme limits: Q$\chi$PT and standard $\chi$PT.  
We have clarified
 that, in the fully quenched limit, how the quenched infrared 
divergences arises through the incomplete cancellation of the pion pole and 
the shifted pole.  
We have also shown that the unquenched limit of PQ$\chi$PT is not, as 
naively expected, standard $\chi$PT, but $\chi$PT with the $\eta'$ meson.  
What we have achieved is an intrapolation between $\chi$PT and Q$\chi$PT, 
which was one of the motivations behind Bernard and Golterman's original 
invention  of partially quenched QCD.  

In our treatment of the vector and tensor meson fields we have adopted the 
heavy particle formalism, which has now became a standard technique in 
the treatment of matter fields in chiral perturbation theory.  
By using the heavy particle formalism one has excluded the effects of 
heavy particle number violating processes like $\rho\to\pi\pi\to\rho$, which 
is not captured in our Lagrangian.  
One may wonder if this would lead to substantial corrections to our 
results.  
Through a model calculation, this issue has been addressed in 
Ref.~\cite{LC}, where it has been found that  
the effect of heavy particle number violating processes is negligibly 
small.  
This problem may be more severe for tensor mesons, but so far no
reliable estimate has been made on these corrections.  
On the other hand, more relevant to the present study, 
there are also $1/M$ correction due to the finite masses 
of the vector and tensor mesons under study.  
The tensor mesons ($M\sim 1.4$ GeV) are probably heavy enough (note that 
$\chi$PT works well for the N--$\Delta$ system with $M\sim 1$ GeV).  
The vector meson are not that heavy at all, and there may be sizable $1/M$ 
corrections, especially, when one has to couple the $\rho$ meson ($m_\rho = 
0.770$ GeV) to the $\eta'$ meson ($m_{\eta'} = 0.958$ GeV).  
It is evident that more studies on these corrections would provide
invaluable information to our understanding of quenched and partially
quenched QCD.  

We thank M.~Alford, C.~Bernard, P.~Lepage, C.~Michael, S.~Sharpe and 
F.~Wilczek for useful discussions and correspondences and C.~Bernard
for careful reading of the manuscript and invaluable comments.

 

\begin{thebibliography}{1}
\bibitem{spec1}
D. Weingarten, Phys. Lett. {\bf 109B} (1982) 57;\\
H. Hamber and G. Parisi, Phys. Rev. Lett. {\bf 47} (1981) 1792.

\bibitem{spec2} T. Bhattacharya, R. Gupta, G. Kilcup and S.R. Sharpe,
Phys. Rev. {\bf D53} (1996) 6486.

\bibitem{spec3} C. Bernard \sl et.al. \rm (MILC Collaboration),
Nucl. Phys. {\bf 53} [Proc. Suppl.] (1997) 212.

\bibitem{spec4} R. Kenway \sl et.al. \rm (UKQCD Collaboration),
Nucl. Phys. {\bf 53} [Proc. Suppl.] (1997) 206.

\bibitem{spec5} S. Aoki \sl et.al. \rm (JLQCD Collaboration), 
Nucl. Phys. {\bf 53} [Proc. Suppl.] (1997) 209.

\bibitem{spec6} U. Glaessner \sl et.al. \rm (SESAM Collaboration),
Nucl. Phys. {\bf 53} [Proc. Suppl.] (1997) 219.

\bibitem{specrev} For a review, see S. Gottlieb, Nucl. Phys. {\bf 53}
[Proc. Suppl.] (1997) 155.

\bibitem{orbitaldata1}
T.~A.~DeGrand and M.~W. Hecht, Phys.~Rev.~{\bf D46} (1992) 3937.

\bibitem{orbitaldata2}
P.~Lacock, C.~Michael, P.~Boyle and P.~Rawland, Phys.~Rev.~{\bf D54}
(1996) 6997; Phys. Lett. {\bf B401} (1997) 308.

\bibitem{orbitaldata3} C.~Bernard \sl et.al.\rm (MILC Collaboration),
\sl Exotic Mesons in Quenched Lattice QCD \rm, \tt hep-lat/9707008 \rm.

\bibitem{politzer} H.D. Politzer, Phys. Lett. {\bf 116B} (1982) 171.

\bibitem{S1} S.R.~Sharpe, Phys.~Rev.~{\bf D41} (1990) 3233.
\bibitem{S2} S.R.~Sharpe, Phys.~Rev.~{\bf D46} (1992) 3146.

\bibitem{BG1} C.W.~Bernard and M.F.~Golterman, Phys.~Rev.~{\bf D46} 
(1992) 853.
\bibitem{BG2} C.W.~Bernard and M.F.~Golterman, Phys.~Rev.~{\bf D49} 
(1994) 486.

\bibitem{gupta} R. Gupta, Nucl. Phys. {\bf 42} [Proc. Suppl.] 
(1994) 85. 

\bibitem{kimsinclair} S. Kim and D.K. Sinclair, Phys. Rev. {\bf D52} 
(1995) 2614.

\bibitem{michael} P. Lacock and C. Michael (UKQCD Collaboration),
Phys. Rev. {\bf D52} (1995) 5213.
 
\bibitem{sharpereview} S.R.~Sharpe, Nucl. Phys. [Proc.Suppl.] {\bf 53}
(1997) 181.

\bibitem{Greview} M.F.~Golterman, \sl Chiral Perturbation Theory and
the Quenched Approximation of QCD, \rm {\tt hep-lat/9411005} .

\bibitem{morel} A. Morel, J. Physique (Paris) {\bf 48} (1987) 111.

\bibitem{pentronzio} G.M. de Divitiis, R. Frezzotti, M. Masetti and
R. Petronzio, Phys. Lett. {\bf 387B} (1996) 829.

\bibitem{sharpe} S.R.~Sharpe,
\sl Enhanced Chiral Logarithms in Partially Quenched QCD, \rm 
\tt hep-lat/9707018. \rm

\bibitem{sharpebaryon} J.N.~Labrenz and S.R.~Sharpe, 
Phys.~Rev.~{\bf D54} (1996) 4595.

\bibitem{sharpeheavy}
M.~Booth, Phys. Rev. {\bf D51} (1995) 2338;\\
S.R.~Sharpe and Y. Zhang, Phys. Rev. {\bf D53} (1996) 5125.

\bibitem{eichten} W.~Bardeen, A.~Dunca, E.~Eichten and H.~Thacker,
\sl Quenched Approximation Artifacts: A Detailed Study in Two-Dimensional
QED, \rm \tt hep-lat/9705002. \rm

\bibitem{F} M.~Booth, G.~Chiladze and A.F.~Falk. Phys.~Rev.~{\bf D55}
(1997) 3092.

\bibitem{GL}
J.~Gasser and H.~Leutwyler, Nucl.~Phys.~{\bf B250} (1985) 465.

\bibitem{JMW} E.~Jenkins, A.V.~Manohar and M.B.~Wise,
Phys.~Rev.~Lett.~{\bf 75} (1995) 2272.

\bibitem{CR} C.K.~Chow and S.J.~Rey, {\sl Chiral Perturbation Theory
for Tensor Mesons}, {\tt hep-ph/9708355}.

\bibitem{reparametrization} 
A.~Falk, M.~Neubert and M.~Luke, Nucl.~Phys.~{\bf B388} (1992) 363.

\bibitem{nrqcd} C.T.H.~Davies \sl et.al. \rm 
(NRQCD Collaboration), Phys.~Rev.~{\bf D50} (1994) 6963;
 Phys. Rev. {\bf D52} 
(1995) 6519; Phys.~Lett.~{\bf B382} (1996) 131.

\bibitem{LC} D.B.~Leinweber and T.D.~Cohen, Phys.~Rev.~{\bf D49} 
(1994) 3512.  

\end{thebibliography}
\end{document}